\documentclass[12pt, journal, onecolumn, final]{IEEEtran}
\usepackage{upgreek}
\usepackage[utf8]{inputenc}
\usepackage[T1]{fontenc}

\usepackage{hyperref}
\usepackage{graphicx}
\usepackage{authblk}
\usepackage{amsmath}

\title{Discrete symmetries tested at 10$^{-4}$ precision using linear polarization of photons from positronium annihilations}

\author[1,2]{\mbox{Pawe\l{} Moskal}}
\author[1,2,*]{\mbox{Eryk Czerwi{\'n}ski}}
\author[1,2]{\mbox{Juhi Raj}}
\author[2,3]{\mbox{Steven~D. Bass}}
\author[1,2]{\mbox{Ermias Y. Beyene}}
\author[1,2]{\mbox{Neha Chug}}
\author[1,2]{\mbox{Aur{\'e}lien Coussat}}
\author[4]{\mbox{Catalina Curceanu}}
\author[1,2]{\mbox{Meysam Dadgar}}
\author[1,2]{\mbox{Manish Das}}
\author[1,2]{\mbox{Kamil Dulski}}
\author[1,2]{\mbox{Aleksander Gajos}}
\author[5]{\mbox{Marek Gorgol}}
\author[6]{\mbox{Beatrix~C. Hiesmayr}}
\author[5]{\mbox{Bo{\.z}ena Jasi{\'n}ska}}
\author[1,2]{\mbox{Krzysztof Kacprzak}}
\author[1,2]{\mbox{Tevfik Kaplanoglu}}
\author[1,2]{\mbox{\L{}ukasz Kap\l{}on}}
\author[7]{\mbox{Konrad Klimaszewski}}
\author[7]{\mbox{Pawe\l{} Konieczka}}
\author[8,2]{\mbox{Grzegorz Korcyl}}
\author[1,2]{\mbox{Tomasz Kozik}}
\author[9]{\mbox{Wojciech Krzemie{\'n}}}
\author[1,2]{\mbox{Deepak Kumar}}
\author[1,2]{\mbox{Simbarashe Moyo}}
\author[1,2]{\mbox{Wiktor Mryka}}
\author[1,2]{\mbox{Szymon Nied{\'z}wiecki}}
\author[1,2]{\mbox{Szymon Parzych}}
\author[1,2]{\mbox{Elena P\'erez~del~R\'{\i}o}}
\author[7]{\mbox{Lech Raczy{\'n}ski}}
\author[1,2]{\mbox{Sushil Sharma}}
\author[1,2]{\mbox{Shivani Choudhary}}
\author[7]{\mbox{Roman~Y. Shopa}}
\author[1,2]{\mbox{Micha\l{} Silarski}}
\author[1,2]{\mbox{Magdalena Skurzok}}
\author[1,2]{\mbox{Ewa~{\L{}}. St\k{e}pie{\'n}}}
\author[1,2]{\mbox{Pooja Tanty}}
\author[1,2]{\mbox{Faranak Tayefi Ardebili}}
\author[1,2]{\mbox{Keyvan Tayefi Ardebili}}
\author[1,2]{\mbox{Kavya Valsan Eliyan}}
\author[7]{\mbox{Wojciech Wi{\'s}licki}}

\affil[1]{Marian Smoluchowski Institute of Physics, Jagiellonian University, Krak\'ow, Poland}
\affil[2]{Centre for Theranostics, Jagiellonian University, Krak\'ow, Poland}
\affil[3]{Kitzb\"uhel Centre for Physics, Kitzb\"uhel, Austria}
\affil[4]{INFN, Laboratori Nazionali di Frascati, Frascati, Italy}
\affil[5]{Institute of Physics, Maria Curie-Sk\l{}odowska University, Lublin, Poland}
\affil[6]{Faculty of Physics, University of Vienna, Vienna, Austria}
\affil[7]{Department of Complex Systems, National Centre for Nuclear Research, Otwock-\'Swierk, Poland}
\affil[8]{Institute of Applied Computer Science, Jagiellonian University, Krak\'ow, Poland}
\affil[9]{High Energy Physics Division, National Centre for Nuclear Research, Otwock-\'Swierk, Poland}
\affil[*]{corresponding author, e-mail: eryk.czerwinski@uj.edu.pl}

\begin{document}
\maketitle
\begin{abstract}Discrete symmetries play an important role in particle physics with violation of CP connected to the matter-antimatter imbalance
in the Universe. 
We report the most precise test of P, T and CP invariance in decays of ortho-positronium, 
performed 
with methodology involving polarization of photons from these decays.
Positronium, the simplest bound state of an electron and positron,
is of recent interest with discrepancies reported between measured hyperfine energy structure and theory at the level of 
$10^{-4}$ 
signaling a need for better understanding of the  positronium system at this level.
We test 
discrete symmetries 
using photon polarizations determined via Compton scattering in the dedicated J-PET tomograph on an event-by-event basis and without the need to control the spin of the positronium with an external magnetic field, 
in contrast to previous experiments.
Our result is consistent with QED expectations at the level of 0.0007 and one standard deviation.
\end{abstract}
%%%%%%%%%%%%%%%%%%%%%%%%%%%%%%%%%%%%%%%%%%%%%%%%
%\IEEEkeywords{CP, positronium, polarization, J-PET}

\section*{Introduction}
Positronium, Ps, is a bound state of an electron and positron with its physics governed by
quantum electrodynamics,  QED.
For describing Ps one commonly 
uses 
non-relativistic 
QED bound state theory.
While 
this approach is mostly successful, 
recent hyperfine structure, HFS, spectroscopy measurements have revealed a 4.5 standard deviations anomaly between experiment and theory 
at the level of one part in $10^4$~\cite{Gurung:2020hms}
prompting new thinking about Ps structure and interactions 
-- for recent discussion see~\cite{Adkins:2022omi,Bass:2023dmv}.

Here, we investigate
properties of the ortho positronium, o-Ps, spin with respect to discrete symmetries.
As a bound state, 
o-Ps should respect the symmetries of its constituents, including discrete symmetries involving parity P, 
charge conjugation C 
and time reversal T invariance~\cite{Bass:2019ibo}.
Fundamental 
QED respects P, C, T symmetries as well as the combinations CP and CPT.
This paper presents the world's most precise test of T, P and CP invariance in o-Ps decays. The measurement is
realized by a 
method 
using 
the polarization of photons from o-Ps decays.

For a single electron or positron, C and CPT
are seen to be 
working to 1 part in $10^{12}$ in their anomalous magnetic moments $a_e=(g-2)/2$~\cite{Fan:2022eto, VanDyck:1987ay}. 
The symmetry between electrons and positrons is also manifested in comparison of their masses 
$
(m_{e^+} - m_{e^-}) / m_{\rm average} < 8 \times 10^{-9} \ $
and electric charges
$\lvert q_{e^+} + q_{e^-} \rvert/ e  < 4 \times 10^{-8}$~\cite{ParticleDataGroup:2022pth}.
 CPT is a general property of relativistic quantum field theories
beyond these charged leptons.
A further recent test
is the 
measurement of the antiproton-to-proton charge–mass ratio resulting in a 16-parts-per-trillion fractional precision in  CPT invariance~\cite{BASE:2022yvh}.

For CP, important information comes from electron electric dipole moment (eEDM).
The tiny value 
$\lvert d_e\rvert < 4.1 \times 10^{-30} e$cm~\cite{Roussy:2022cmp}
(see also~\cite{Andreev:2018ayy,Hudson:2011zz,Cairncross:2017fip})
constrains the scale of 
any new  CP violating interactions coupling
to the electron.
If such interactions
couple with similar strength to Standard Model particles, 
then one finds constraints on the
heavy particle masses
similar
to the constraints from the Large Hadron Collider at CERN.
If, instead, 
the new interactions should involve 
ultra-light particles, 
then one finds that
their couplings to the electron should be less than about  
$\alpha_{\rm eff} \sim 5 \times 10^{-9}$~\cite{Bass:2019ibo}.
Some new  CP violation from beyond the Standard Model
is needed to explain baryogenesis~\cite{Sakharov:1967dj}
- hence 
the interest in looking for such couplings. 
There are hints for possible CP violation in the neutrino sector though conservation is still allowed at the level of 1-2~$\sigma$~\cite{ParticleDataGroup:2022pth,T2K:2021xwb,NOvA:2021nfi}.

Based on the EDM 
constraints 
one
expects CP to be working in Ps decays down to branching ratios at least about $10^{-9}$~\cite{Bass:2019ibo}.
This 
has been explored in studies of CP-odd correlations~\cite{Sozzi:2008zza}, e.g.,
between final state photon momenta and the spin
of the Ps.
These experiments used ortho-positronium
which decays into three photons with a lifetime in vacuum of 142~ns~\cite{Vallery:2003iz}, and measured the correlation
\begin{equation}
O_1 =
( \mathbf{S} \cdot \mathbf{k}_1 ) ( \mathbf{S} \cdot ( \mathbf{k}_1 \times \mathbf{k}_2 ) 
\label{eq:acp}
\end{equation}
with $\mathbf{S}$ the o-Ps spin vector and $\mathbf{k}_i$ the momenta of the emitted photons 
defined with magnitude $\mathrm{k}_1 > \mathrm{k}_2 > \mathrm{k}_3$, 
and found the result 
$
\langle O_1 \rangle =  
0.0013 \pm 0.0022
$~\cite{Yamazaki:2009hp}
-- consistent with zero at the level of $2\times10^{-3}$.
Since Ps freely decays in vacuum to massless photons,  it is not an eigenstate of T.
This means that one can get
 CP,  T and CPT violation mimicking final state interactions with magnitudes only detectable at the prevision level of about $10^{-9}$-$10^{-10}$~\cite{Bernreuther:1988tt},
beyond the scope of the present experiments.

In this paper
we develop a methodology made possible using the \mbox{J-PET} tomograph in Krak\'ow using polarizations of the emitted photons, which are determined from Compton rescattering in the detector~\cite{Moskal:2016moj}.
No magnetic field to control the o-Ps spin is needed in the experiment.
The maximal cross section of the Compton scattering is for the direction perpendicular to the electric field and polarization axis $\mathbf{\upepsilon}$ of the incident photon~\cite{Klein:1929ab,Moskal:2018pus}.
This leads to defining the polarization related quantities 
\begin{equation}
\mathbf{\upepsilon}_i = \mathbf{k}_i \times \mathbf{k'}_i~/~\lvert~\mathbf{k}_i \times \mathbf{k'}_i~\rvert,
\label{eq:epsilon}
\end{equation}
where $\mathbf{k}_i$ and $\mathbf{k'}_i$ are the momenta of a photon from the positronium decay 
before and after
Compton scattering in the detector, respectively~\cite{Moskal:2016moj}.
These $\mathbf{\upepsilon}_i$ vectors are most likely to be along the axis of the incident photon polarization vector
and are even under  P and T transformations.

One may then 
consider new correlations. 
Taking the polarization vector of one of photons $ \mathbf{\upepsilon}_i$ and momentum vector of another photon $\mathbf{k}_j$,
we construct the 
momentum-polarization
correlations~\cite{Moskal:2016moj}
\begin{equation}
O_2 =
\mathbf{\upepsilon}_i \cdot \mathbf{k}_j
=
\text{cos}(\omega_{ij})
\label{eq:operator}
\end{equation}
for all three independent combinations of these vectors, 
$(i,j)=(1,2), (1,3), (2,3)$
with 
$\omega_{ij}$ being
the angle 
between 
the 
$ \mathbf{\upepsilon}_i$ 
and $\mathbf{k}_j$ vectors.
This correlation
$O_2$ is odd 
under P, T and CP 
transformations.
If the expectation value of $O_2$ does not vanish, then each of T, P, and CP symmetries would be violated in the 
o-Ps decay.
Measurement of the correlation $O_2$
can be performed
without an external magnetic field and without control of the o-Ps spin.

Here we present an investigation of discrete symmetries in the o-Ps system
based on 
the momenta and
polarizations of the emitted photons Eq.~(\ref{eq:operator})
in o-Ps decays
over the entire range of $\omega_{ij}$:
\begin{equation}
\langle O_2 \rangle
= \langle \text{cos}(\omega_{ij}) \rangle
= \langle \mathbf{\upepsilon}_i \cdot \mathbf{k}_j~/~\mathrm{k}_j \rangle.
\label{eq:alpha}
\end{equation}
We calculate $\langle O_2 \rangle$ from a distribution which is the
sum of all independent combinations of $\omega_{ij}$.
We find a value consistent with zero
at 68\% confidence level, as expected from the underlying QED 
with a threefold precision improvement over the
previous measurements 
of the  CP-odd correlation,
Eq.(\ref{eq:acp}),
where the 
o-Ps spin was used to define the
correlation.
The bound state o-Ps decays obeys the CP 
symmetry of the underlying QED dynamics. 
 Here one is probing the discrete symmetry properties of QED. Weak interaction effects are characterized by a factor $G_F m_e^2 \approx 10^{-11}$ with $G_F$ the Fermi constant, and would only be manifested with very much enhanced precision.

%%%%%%%%%%%%%%%%%%%%%%%%%%%%%%%%%%%%%%%%%%%%%%%%%%%%%%%%%%%%%%%%%%%%%
\section*{Results}
\subsection*{Detector}
\begin{figure}[t]
 \centering
  \includegraphics[width=\textwidth]{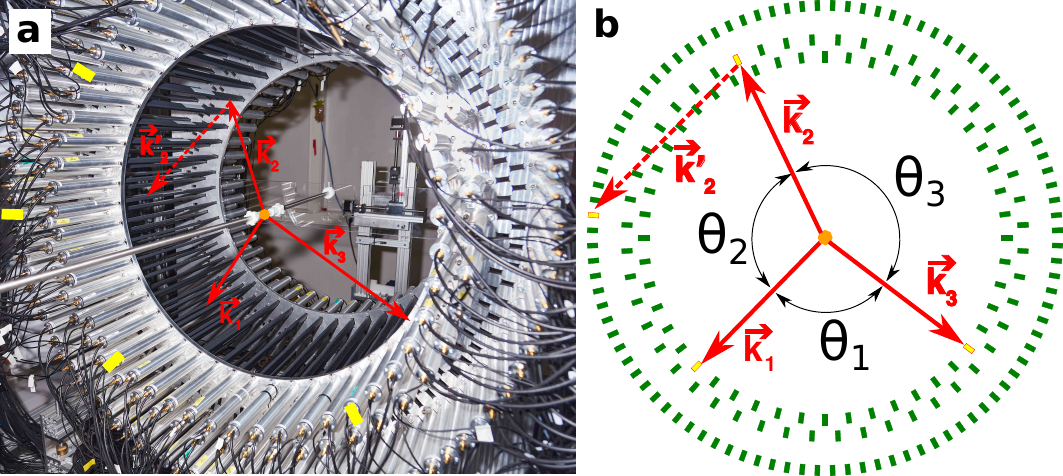}
  \caption{The J-PET detection system.
The orange dot indicates the position of the sodium source.
The superimposed solid red arrows indicate momenta of annihilation photons ($\mathrm{k}_{1} > \mathrm{k}_{2} > \mathrm{k}_{3}$)
originating from the decay of ortho-Positronium.
The dashed red vector represents the momentum of the secondary scattered
photon 
($\mathbf{k'}_{2}$).
Photomultipliers registering signals from these four photons are marked with yellow rectangles.
\textbf{a}~Photograph of the J-PET detector with 
the annihilation chamber installed at the center.
Strips of plastic scintillator wrapped in black foil are mounted between two aluminum plates.
Photomultipliers reading optical signals from these strips are inserted in aluminum tubes with mu-metal insert for optic and magnetic isolation.
\textbf{b}~Scheme of the J-PET detector where scintillators are drawn as green rectangles.
For every selected event the directions of the momentum vectors for the  three annihilation photons
are reconstructed between the known position of radioactive source and the reconstructed hit point.
Due to momentum conservation these three vectors are co-planar (annihilation plane).
In the presented example the photon with medium energy ($\mathbf{k}_{2}$) interacts with the detector material and scatters
as $\mathbf{k'}_{2}$ (forming the scattering plane).
The angles between photon momenta are indicated such as $\theta_1 < \theta_2 < \theta_3$.
Note that ordering of these angles is not directly related to the ordering of the momenta.
}
\label{fig:detector}
\end{figure}

The strategy we use here is to study the discrete symmetries associated with the operator correlation, Eq.~(\ref{eq:operator}), involving the momenta of photons from the o-Ps decay and the 
photon polarization related vectors
$\mathbf{\upepsilon_i}$, which are measured using the Jagiellonian Positron Emission Tomograph 
(J-PET)~\cite{Moskal:2014sra,Niedzwiecki:2017nka,Moskal:sciadv,Moskal:2018pus}.
The J-PET detector is based on plastic scintillators and is designed for total body scanning~\cite{Moskal:2021jwc} in medicine~\cite{Moskal:2018wfc,Moskal:sciadv,Moskal:petclinics} as well as biomedical studies~\cite{Moskal:biorxiv,NRP:2019} and fundamental physics research~\cite{Moskal:2016moj,Bass:2019ibo,Moskal:2018pus}. The J-PET detector is described in more detail in the Methods section.
For the measurement reported here the positrons are emitted from a radioactive $^{22}$Na source placed at the center of the detector (Fig.~\ref{fig:detector}a).

The source is coated with porous polymer material to increase the probability of o-Ps creation~\cite{gorgol} and inserted in the small vacuum chamber to decrease the background contribution from positron annihilation in the air.
The annihilation photons from o-Ps${}\to3\gamma$ are registered in three layers of scintillator strips forming a barrel shaped detector 
(Fig.~\ref{fig:detector}a).
Detection of annihilation photons in a given scintillator is based on registration of a 
light pulse at both ends of the scintillating strip.
The light is collected by means of attached photomultipliers and the interaction, later on referred to as hit, is confirmed if signals at both ends of the strip are over a 
30~mV threshold within a coincidence time 
of 6~ns. 
Fig.~\ref{fig:detector}b presents an example of a signal event of 
o-Ps${}\to3\gamma$ annihilation for the
CP symmetry test.
The novelty of the reported measurement is in the determination of polarization plane of annihilation photons and the experimental coverage of the whole angular range of the tested correlation.
Additionally, application of data acquisition system based on a fast, trigger-less, field-programmable gate array (FPGA)~\cite{Palka:2017wms,FPGA1} and good timing properties of the plastic scintillators used in the experiment~\cite{EJ230,Kaplon:2020tqw,Kaplon:inpress} (short light signals with 1~ns rising and 2~ns falling edges) and a high activity $\beta^+$ radioactive source allowed us to register the highest number of 
o-Ps${}\to3\gamma$ annihilations for discrete symmetry studies so far recorded.

Additionally, the achieved high data throughput make it possible to use high granularity of the active detector elements.
As a consequence the angular resolution in the plane perpendicular to the detector axis is 0.5 degrees, which is important for the determination of the momentum direction.
%%%%%%%%%%%%%%%%%%%%%%%%%%%%%%%%%%%%%%%%%%%%%%%%%%%%%%%%%%%%%%%%%%%%%
\subsection*{Signal and background}
In order to construct the operator correlation defined in Eq.~(\ref{eq:operator}) three vectors of photon momenta are required:
the momentum vectors 
of a Compton scattered photon before and after scattering, 
and an arbitrarily chosen 
one of two remaining photons from the o-Ps annihilation.
However, for proper \mbox{o-Ps}${}\to3\gamma$ event identification and $\mathrm{k}_{1} > \mathrm{k}_{2} > \mathrm{k}_{3}$ ordering,
the momenta of all photons from o-Ps decay must be reconstructed.
In this work we consider the expectation value of the distribution of the sum of three independent operators constructed with the aforementioned vectors.
The momentum vector of a photon is reconstructed on the basis of its origin point and point of interaction with the scintillator.
The origin point is common for photons emitted from 
the o-Ps decay and is equivalent to the source position.
The point and time of interaction with the scintillator are calculated on the basis of the difference and sum of times,
respectively, of registered signals at both ends of scintillator strips~\cite{Moskal:2014sra}.
The main 
experimental background 
to o-Ps${}\to3\gamma$ signal events (described in detail in the Methods section) consists of
(i) p-Ps${}\to2\gamma$ events with single scattering registered,
(ii) events with multiple scattering of single photon between active elements of detector 
and (iii) cosmic rays.
%%%%%%%%%%%%%%%%%%%%%%%%%%%%%%%%%%%%%%%%%%%%%%%%%%%%%%%%%%%%%%%%%%%%%
\subsection*{Analysis scheme}
A signal event consists of four  depositions of energy inside scintillating strips: three from o-Ps${}\to3\gamma$ and one from registered Compton scattering.
Hits with energy deposition of at least 31~keV are registered by the data acquisition system, DAQ.
Background interactions from p-Ps${}\to2\gamma$ and cosmic radiation with high energy depositions are suppressed by the requirement of 
time-over-threshold, 
TOT, less than 17~ns.
Hits from o-Ps${}\to3\gamma$ decays (for each combination of three hits within an event) are identified based on  
the angular correlation between annihilation photons,
comparison of their emission time
and coplanarity of 
the momentum vectors of the 
annihilation photons.
The energy of the annihilation photons
is calculated from the angular dependence between all three
photons from the o-Ps decay~\cite{Kaminska:2016fsn} 
and the momenta are ordered as $\mathrm{k}_1 > \mathrm{k}_2 > \mathrm{k}_3$.
Finally, the assignment of one of the remaining hits in the event to one of 
the photons
originating from o-Ps${}\to3\gamma$
is based on the smallest value of difference between calculated and measured
time of flight of photon between $\mathbf{k}_i$ and $\mathbf{{k'}_i}$ interactions.
The detailed description of the applied selection criteria is in Methods section.
After the aforementioned selection the final sample consists of $7.7\times10^5$ events.
The angular correlation between momenta of annihilation photons for the final sample is presented in the Fig.~\ref{fig:result}a.
%%%%%%%%%%%%%%%%%%%%%%%%%%%%%%%%%%%%%%%%%%%%%%%%%%%%%%%%%%%%%%%%%%%%%
\subsection*{Expectation value of the correlation $O_2$}
\begin{figure}[t]
\centering
\includegraphics[width=\textwidth]{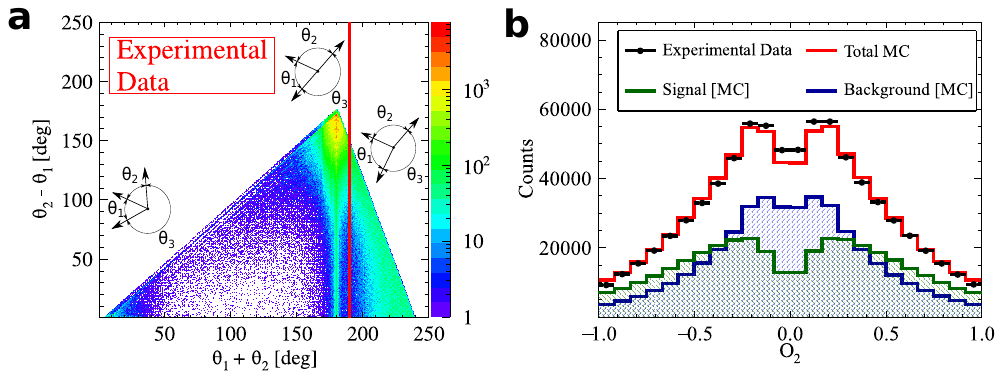}
 \caption{
Composition of the experimental data sample.
\textbf{a}~Distribution of the sum and difference of the two smallest angles between photon
momenta ($\theta_1 < \theta_2 < \theta_3$).
The superimposed black pictographs present three different orientations of the momentum vectors for
multiple scattered events (bottom left region), p-Ps${}\to2\gamma$ events with single scattering
(vertical band around $\theta_{1}+\theta_{2}=180^\circ$) and o-Ps${}\to3\gamma$ signal events (bottom right region).
The red vertical line corresponds to a $\theta_{1}+\theta_{2}\geq190^\circ$ cut applied for the signal selection.
\textbf{b}~Measured distribution of asymmetry operator Eq.~(\ref{eq:alpha}) for experimental data (black circles) and simulated histograms for signal (green),
background (blue) and combined signal and background (red). 
The discrepancy between simulated distribution and data points for the two central bins may be explained by the rapid change of efficiency distribution
in that region, but this effect is negligible comparing to the achieved accuracy of the final result.
}
\label{fig:result}
\end{figure}

The distribution of the  reconstructed correlation defined in 
Eq.~(\ref{eq:operator}) is presented in the Fig.~\ref{fig:result}b.
For the first time the whole range of the CP asymmetry operator is measured.
For the distribution of $O_2$ operator
the background expected on the grounds of performed Monte Carlo simulations is subtracted from the experimental distribution.
The resulting distribution is corrected for the detector acceptance and analysis efficiency.
The 
expectation value of the 
operator correlation
$O_2$ is determined to be
\begin{equation}
\begin{aligned}
 \langle O_2 \rangle=0.0005 \pm 0.0007_{\text{stat.}}.
  \end{aligned}
\end{equation}
The systematic error contributions to this result are estimated from hit spatial, temporal and energy resolutions.
The possible influence of cosmic rays is tested on the basis of a dedicated measurement without the positron source, but with an
identical data processing scheme to that used  for 
the $\langle O_2 \rangle$ 
determination.
No significant systematic error from any contribution is found.
The expectation value of the 
operator $O_2$ is consistet with zero within achieved accuracy, therefore no P, T and CP asymmetry is observed.
%%%%%%%%%%%%%%%%%%%%%%%%%%%%%%%%%%%%%%%%%%%%%%%%%%%%%%%%%%%%%%%%%%%%%
\section*{Discussion}
Our methodology using the polarization of photons from positronium decays 
has allowed us to make the 
world's presently most accurate test of CP symmetry in o-Ps decays.
The experiment uses 
the polarizations of photons emitted in the decay measured through the
non-local correlation
in Eq.~(\ref{eq:operator}),
which 
is independent of the o-Ps spin.
It involves the o-Ps decay and rescattering in the detector. 
Previous tests of CP-odd~\cite{Yamazaki:2009hp} and CPT-odd~\cite{cpt-vetter} decays of o-Ps
were conducted 
by measuring 
angular correlations between momenta of the annihilation photons
and the spin of the ortho-positronium
only at specific fixed angles.
Recently the J-PET group improved the test of CPT symmetry 
by measuring 
a momentum-spin correlation 
with full angular coverage~\cite{cpt-jpet}.
The CP
result reported here is also obtained using the 
full kinematic range 
of photons appearing in the correlation
$O_2$.
With our method 
the CP test is performed 
without the need 
to
control the o-Ps spin using an external magnetic field. 
It is the first simultaneous test of P, T and CP symmetries using the
angular correlation
between momentum of one of the annihilation
photons and the polarization plane of another annihilation photon.
\begin{figure}[b]
\centering
\includegraphics[width=0.7\textwidth]{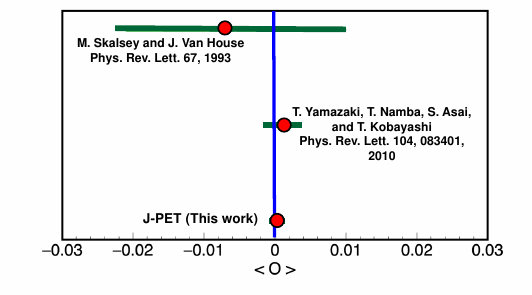}
\caption{
Summary of searches for CP-odd ortho-Positronium decays.
The two upper results~\cite{Skalsey:1991vt,Yamazaki:2009hp} are performed for the operator correlation $O_1$ defined
in Eq.~(\ref{eq:acp}), whereas J-PET is using the new correlation 
$O_2$
constructed with the  polarization vector in Eq.~(\ref{eq:operator}).
The blue vertical line indicates no CP symmetry violation, while the green bars for each measurement correspond to the total uncertainty
calculated as statistical and systematic uncertainties combined in quadrature.
}
\label{fig:cpcompare}
\end{figure}
The result reached the  
precision of
${\cal O}(10^{-4})$, 
which represents a threefold 
improvement 
in the search for CP-odd decays of o-Ps~(Fig.~\ref{fig:cpcompare}).
The use of polarization of photons for correlations like 
$\mathbf{\upepsilon} \cdot \mathbf{k}$ or $\mathbf{\upepsilon} \cdot \mathbf{S}$, where $\mathbf{S}$ is the spin of the positronium, 
opens a new class of discrete symmetry tests in positronium decays~\cite{Moskal:2016moj}.

The new result might be further improved 
using the methods introduced here 
together with upgrades in  
the J-PET detector. 
These experiments will be conducted with a modular 
J-PET detector having about 20 times higher sensitivity for the registration of ortho-positronium.
The modular version of the J-PET system~\cite{Eliyan} with increased acceptance is currently being used for a measurement of the P, T, CP and CPT symmetries
with a goal of reaching $10^{-5}$ accuracy.
%%%%%%%%%%%%%%%%%%%%%%%%%%%%%%%%%%%%%%%%%%%%%%%%%here go methods
\section*{Methods}
%%%%%%%%%%%%%%%%%%%%%%%
\subsection*{Experimental setup}
\begin{figure}[t]
\centering
  \includegraphics[width=0.7\textwidth]{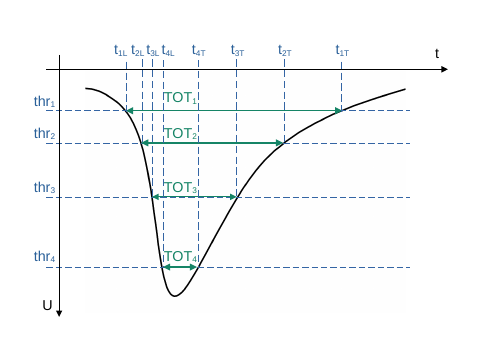}
  \caption{The idea of a
  TOT measurement for constant thresholds $thr_i$,
where i=1,2,3,4. A negative electric signal (black line) is probed in the voltage domain $U$ at four voltage thresholds allowing for
determination of crossing time $t$, with leading $t_{iL}$ and trailing $t_{iT}$ edge of the signal.
The energy carried by the signal is therefore proportional to the area under the signal,
which is estimated as a sum of areas of rectangles limited by neighboring thresholds and registered times.
The energy deposited by a photon 
is proportional to the light collected by photomultipliers at both ends of scintillator strip.
The TOT for a given deposition (hit)~\cite{Sharma:2023,Sharma:2019zwo,Palka:2017wms} is calculated as
the normalized sum of products of a difference of consecutive thresholds with respect to the baseline ${\rm thr_0}=0$~mV
and TOT measurements at both ends of the scintillator strip for each threshold. The 
normalization factor is the  difference between two highest thresholds, namely
${\rm TOT= \sum_{j=1}^{2} \sum_{i=1}^{4} TOT_{i}^{j} \cdot (thr_i-thr_{i-1})/(thr_4-thr_3)}$,
where
$j$ counts TOT measurements at both ends of the scintillator strip.
}
\label{fig:TOT}
\end{figure}

The J-PET shown in Fig~\ref{fig:detector}, 
is a multi-purpose, axially symmetric detector in the form of a barrel
constructed with three layers of plastic scintillator 
strips~\cite{Moskal:2014sra,Czerwinski:2017ibo,Dulski:2020pqi}.
Two inner layers consist of 48 strips each,
placed at 425~mm and 467.5~mm radius, respectively,
with the second layer rotated by 3.75$^\circ$ with respect to
the first one.
The outer layer is composed of 96 strips at radius 575~mm.
A single strip of J-PET is $500\times19\times7$~mm$^3$
and made of fast timing plastic scintillator~\cite{Moskal:2014sra,EJ230,Czerwinski:2017ibo,Dulski:2020pqi}
wrapped with two kinds of foils: external 
(for optical isolation) and an internal one to reflect the light from the scintillator.
The position of a photon interaction along a scintillator strip
is derived from the time difference of signals
from two photomultipliers attached to a given strip,
whereas the time of interaction is calculated from a sum of times of these signals.
Each $19\times7$~mm$^2$ side is optically connected to the R9800 Hamamatsu photomultiplier~\cite{Moskal:2014sra,Niedzwiecki:2017nka}.
Signals from 192 photomultipliers are probed in voltage domain
at 4 different amplitude thresholds~(Fig.~\ref{fig:TOT}).
In total up to 8 measurements of time
(leading and trailing edges) are performed
allowing for precise signal start time derivation
and time-over-threshold
measurement
equivalent to the photon deposited energy determination~\cite{Sharma:2023,Sharma:2019zwo,Palka:2017wms}.
In the reported measurement the amplitude thresholds are
equal to 30, 80, 190 and 300~mV.
All Time-to-Digital Converter (TDC) channels are distributed on the eight
Trigger Readout Boards (TRBs) in the trigger-less manner~\cite{FPGA1}.
Large amounts of data are  registered due to trigger-less data acquisition system DAQ~\cite{FPGA1},
namely for the 1~MBq source there are $10^5$ hits per second collected, which translates into 28~MBps of recorded data.
The Lempel-Ziv-Markov chain algorithm~\cite{lzma} is used to compress the data.
The 122 days of data taking reported here resulted in 100~TB of archived data.
For long term storage, data were recorded on magnetic tapes in Linear Tape-Open version 7 (LTO-7).
Data analysis of J-PET files was performed with a dedicated analysis framework~\cite{Framework,Framework2}.
In the reported measurement four data campaigns were carried out:
two with $^{22}$Na source of 5~MBq activity and two with activity of 1~MBq.
The source was inserted between two 3~mm thick pads of \mbox{XAD-4} porous polymer~\cite{Jasinska:2016qsf} and placed in the center of PA6 polyamide cylindrical
chamber of inner diameter of 10~mm located on the axis of the J-PET detector 
(Fig.~\ref{fig:detector}a).
A vacuum system connected to the source holder ensured a pressure at a level of $\sim 1.5\times10^{-4}$~Pa inside its volume.
Taking into account the density of
the chamber material (1.14 g/cm$^3$), the thickness of the outer wall (1 mm) and mass attenuation coefficient~\cite{attcoef},
the attenuation of 
photons
from o-Ps annihilation is estimated to 1\%.

\begin{figure}[b]
\centering
\includegraphics[width=0.7\textwidth]{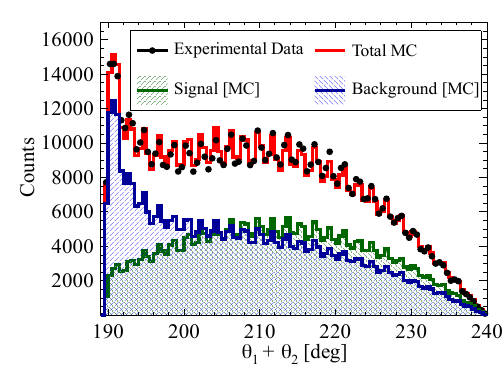}
\caption{
Distribution of the sum of the two smallest relative azimuthal angles~($\theta_{i}$ and $\theta_{j}$) between the registered annihilation photons
(projection on the horizontal axis of experimental data from Fig.~\ref{fig:result}a and simulation
of background and signal events (Fig.~\ref{fig:3D}b and Fig.~\ref{fig:3D}c, respectively).
Experimental data points are marked with black circles, while histograms represent the results of reconstructed Monte Carlo simulations for signal (green), background (blue), and combined signal and background simulated distributions (red).
The experimental histogram contains all the events after the analysis.
Visible multiple maxima and minima are due to distances between scintillating strips (Fig.~\ref{fig:detector}b).
The main maximum at $190^\circ$ is in fact a remaining tail of background component of 
the p-Ps${}\to2\gamma$ process where $\theta_{i}+\theta_{j}=180^\circ$ and
one of
the photons
undergoes a single scattering.
}
\label{fig:normalization}
\end{figure}

For Monte Carlo simulations the geometry and material of the annihilation chamber and active detector elements (scintillator strips)
are implemented in the GEANT4 toolkit~\cite{GEANT4:2002zbu}. Experimental resolution of the whole experimental setup is introduced as Gaussian smearing with standard deviation $\sigma$. For the deposited energy $\sigma_E$=14~keV, for time of the hit $\sigma_T$=225~ps and for Z-position of 
the photon interaction $\sigma_Z$=2.4~cm.
The values of the above mentioned smearing parameters are obtained from the fit of Monte Carlo distributions to data points 
shown in 
Fig.~\ref{fig:normalization}.

The X and Y coordinates of 
photon
interactions are generalized to the center of a given scintillator strip.
In order to reduce the statistical fluctuation of simulated samples of events the generated signal and background events are 3.5 and 2.4 times bigger than contributions found in the experimental data, respectively.
%%%%%%%%%%%%%%%%%%%%%%%
\subsection*{Signal candidates selection}
A signal event is an 
o-Ps${}\to3\gamma$ decay with one of the annihilation 
photons
undergoing Compton scattering.
Therefore a signal candidate consists of four registered 
hits of photons
in scintillator strips:
three coming directly from  annihilation of o-Ps and one as a secondary scattered photon.
\begin{figure}[b]
\centering
  \includegraphics[width=0.7\textwidth]{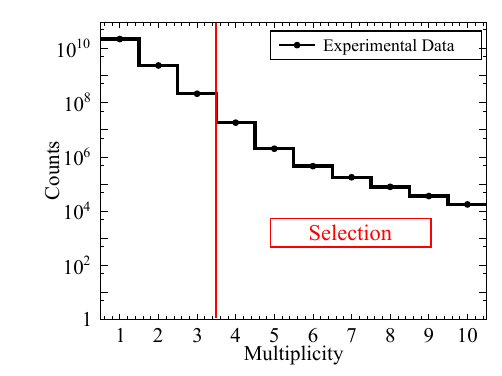}
  \caption{
  Exemplary distribution of hit multiplicity within events.
}
\label{fig:multiplicity}
\end{figure}

\begin{figure}[b]
\centering
\includegraphics[width=\textwidth]{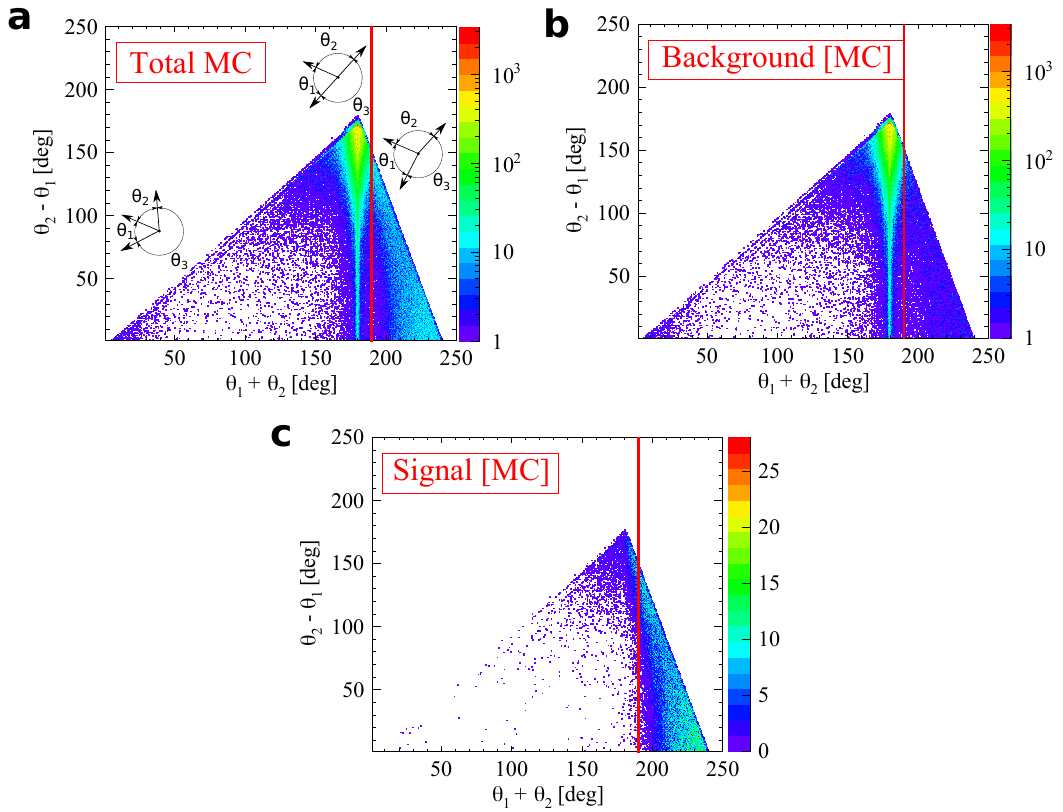}
\caption{
Identification of background to o-Ps${}\to3\gamma$ signal events.
$\theta_1$ and $\theta_2$ indicate two smallest relative angles between momentum vectors of photons.
The superimposed black pictographs at the first plot present three different orientation of momentum vectors for events with multiple scatterings
(bottom left region of each plot), p-Ps${}\to2\gamma$ events with a single scattering
(vertical band around $\theta_{1}+\theta_{2}=180^\circ$) and o-Ps${}\to3\gamma$ signal events (bottom right region).
The red vertical line corresponds to a $\theta_{1}+\theta_{2}\geq190^\circ$ cut applied for signal selection.
\textbf{a}~Full sample of MC simulated events.
\textbf{b}~Background events within the Monte Carlo  sample.
\textbf{c}~Simulated signal events.
}
\label{fig:3D}
\end{figure}

Selection of signal candidates is a three step  process:
\begin{enumerate}
    \item Selection of at least 4 candidate hits within one event (Fig.~\ref{fig:multiplicity}), where each hit fulfills the following conditions:
    \begin{itemize}
        \item the energy deposited in the scintillator must not be smaller than 31~keV to reject multiple scattered hits (this value corresponds in fact to the lowest threshold set at DAQ);
        \item the position of the hit at the scintillator strip must be within $\pm23$~cm window around the center point to suppress hits
	at the ends of scintillators due to scatterings from aluminum plates holding scintillator strips (Fig.~\ref{fig:detector}a);
        \item the registered TOT value must be $\leq 17$~ns to reject hits originating from cosmic radiation (tested with separate data taking campaign without the radioactive source) and to reduce the p-Ps${}\to2\gamma$ background component (Fig.~\ref{fig:cuts}a), as well as the deexitation photon from ${}^{22}$Ne$^*$;
    \end{itemize}

\item
\begin{figure}[b]
\centering
  \includegraphics[width=\textwidth]{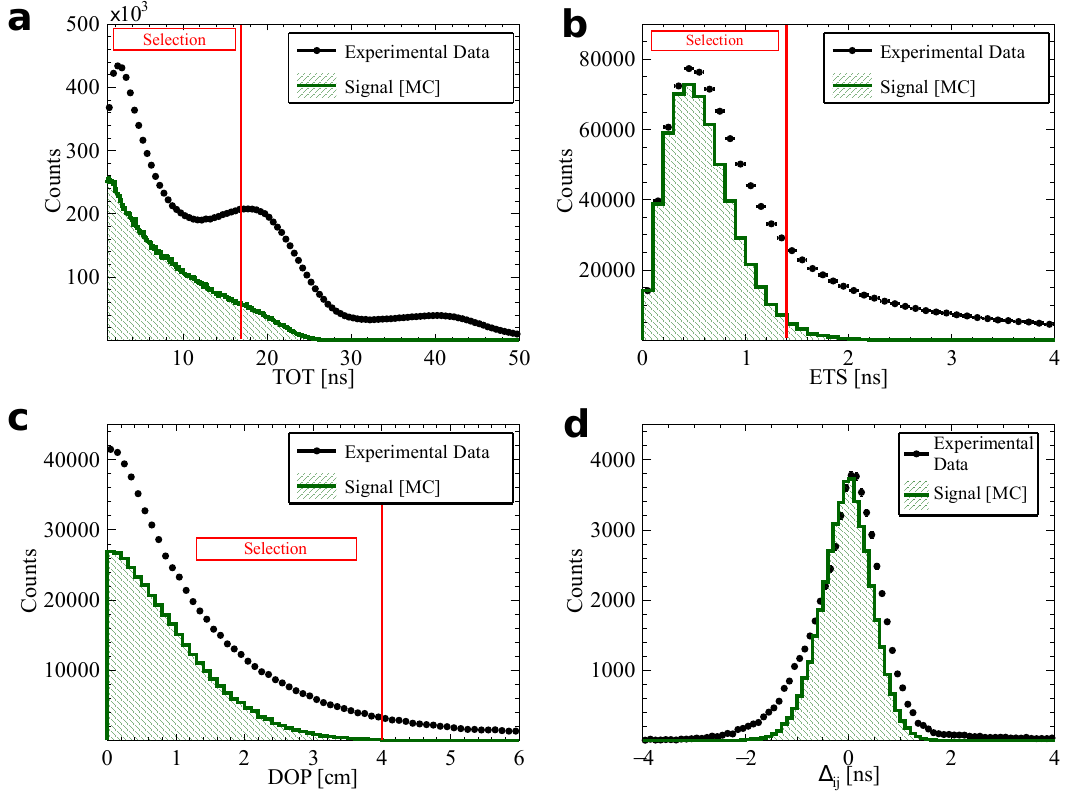}
  \caption{Exemplary spectra for the selection criteria with superimposed red line of the applied cut value.
The definition of each variable is given in the text.
  \textbf{a} Time over threshold TOT for each hit for events with multiplicity greater than 4.
The o-Ps${}\to3\gamma$ candidate is constructed out of 3 hits with the smallest
value of $({\rm ETS})^2 + ({\rm DOP})^2$.
The final candidates are selected after surviving cuts on
ETS (\textbf{b}) and DOP (\textbf{c}).
\textbf{d} The assignment of scattered photon to one of o-Ps${}\to3\gamma$ candidates
is based on the smallest scatter test value.
}
\label{fig:cuts}
\end{figure}
The identification of hits from o-Ps${}\to3\gamma$ decay
was performed as follows:
    \begin{itemize}
        \item the emission time was calculated for each hit as a difference between the registered time (hit time) and a travel time (ratio of distance between source and hit position and speed of light); the emission time spread (ETS) was calculated as a difference between last and first emission time of three candidates;
this ETS must be less than or equal to 1.4~ns to ensure that hits
originate from the same o-Ps decay (Fig.~\ref{fig:cuts}b);
%        \item for a source position of $(s_x,s_y,s_z)$ a distance between annihilation plane (spanned by the annihilation photons' momenta and defined as\linebreak $Ax+By+Cz+D=0$) and the source was calculated as
        \item for a source position of $(s_x,s_y,s_z)$ a distance between annihilation plane (spanned by the annihilation photons' momenta and defined as $Ax+By+Cz+D=0$) and the source was calculated as
\linebreak 
$\textrm{DOP}=\vert A\cdot s_x + B\cdot s_y + C\cdot s_z + D \vert\cdot (A^2 + B^2 + C^2)^{-\frac{1}{2}}$;
the DOP constructed with three candidate hits must be less than or equal to 4~cm to reject hits from multiple scatterings 
(Fig.~\ref{fig:cuts}c);
        \item at the decay plane the sum of the two smallest angles between photon momentum vectors from o-Ps${}\to3\gamma$ decays must be greater than or equal to $190^\circ$
(Fig.~\ref{fig:result}, \ref{fig:normalization} and \ref{fig:3D}) to reject main contribution from p-Ps${}\to2\gamma$ events with multiple scattered photons; 
        \item for events with more than three hits a combination with the smallest $({\rm ETS})^2+({\rm DOP})^2$ value was selected;
    \end{itemize}
    \item After the above mentioned selection of three photons from o-Ps${}\to3\gamma$ decay the 
    assignment of one of the remaining hits in the event as the interaction of a scattered photon from the o-Ps${}\to3\gamma$ decay
was based on the smallest time difference $\Delta_{ij}=(t_j-t_i)-\lvert~\mathbf{r}_j - \mathbf{r}_i~\rvert/c$ between the reconstructed and expected time of flight of
the scattered photon, where the measured time and position of interactions are $t_i,\mathbf{r}_i$ for the
i-th selected annihilation 
photon
and $t_j,\mathbf{r}_j$ for j-th candidate for  
scattering of the i-th photon, respectively 
($i=1,2,3$ and $j=4,..., \textrm{~multiplicity}$), where multiplicity is the number of registered hits per event,
see Fig.~\ref{fig:cuts}d.
\end{enumerate}

\begin{figure}[!p]
\centering
\includegraphics[width=\textwidth]{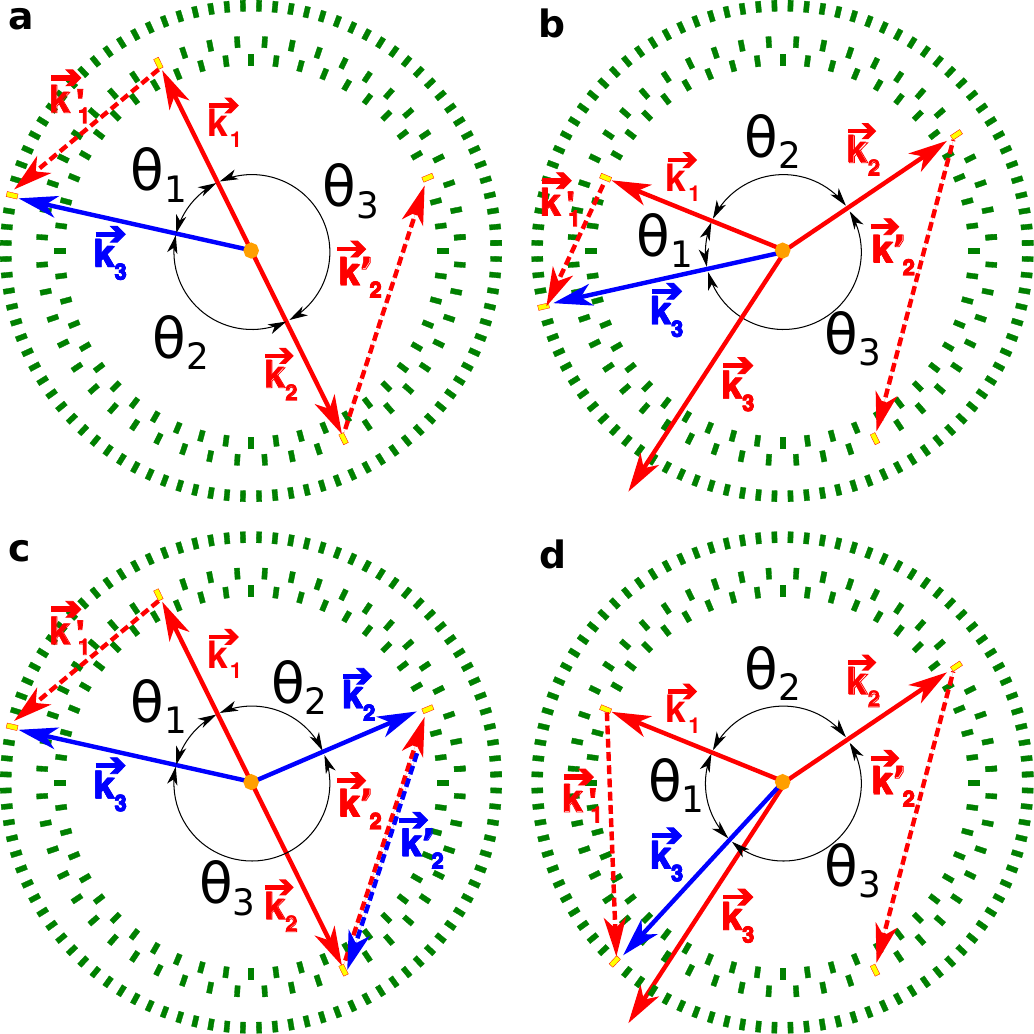}
\caption{
Topology of background events. Scintillators of J-PET are schematically presented as green rectangles.
Scintillators registering the 
photons in presented events are indicated as yellow rectangles.
Solid lines denote 
photons
originating from Ps annihilation, while dashed lines represent scattered 
photons.
Momenta of signal photons are drawn with red color, while incorrectly reconstructed ones - with blue.
The following convention is used:
$\theta_1 < \theta_2 < \theta_3$.
\textbf{a}~An exemplary event of p-Ps${}\to2\gamma$ decay with registration of both annihilation 
photons and two scatterings.
The one with direction of $\mathbf{k'}_{2}$ is correctly recognized during analysis as a scattered hit, while the scattered $\mathbf{k'}_{1}$
is wrongly assigned as $\mathbf{k}_3$.
\textbf{b}~A misreconstructed o-Ps${}\to3\gamma$ decay due to wrong assignment of the scattered hit.
The photon
from o-Ps annihilation marked as $\mathbf{k}_3$ (red) is not detected,
while $\mathbf{k}_2$ scatters as $\mathbf{k'}_{2}$ and is properly reconstructed.
The annihilation
photon
with momentum direction marked as $\mathbf{k}_1$ also scatters. It is not reconstructed as $\mathbf{k'}_{1}$,
but incorrectly reconstructed as annihilation 
photon
$\mathbf{k}_3$ (blue).
Both events presented in  the top row would be rejected by the $\theta_{1}+\theta_{2}\geq190^\circ$ criterion,
while events from the bottom row would be incorrectly accepted.
\textbf{c}~p-Ps${}\to2\gamma$ decay with registration of both annihilation 
photons and two scatterings, but only one of the hits is correctly assigned ($\mathbf{k}_1$).
$\mathbf{k'}_{1}$ is misidentified as $\mathbf{k}_3$ (blue), while $\mathbf{k'}_{2}$ (red) is misidentified as $\mathbf{k}_{2}$ (blue) and $\mathbf{k}_{2}$ (red) as $\mathbf{k'}_{2}$ (blue).
\textbf{d}~An event similar to the one presented in the top right panel, but with a topology immune to the $\theta_{1}+\theta_{2}\geq190^\circ$ cut.
}
\label{fig:bcg}
\end{figure}

The main background contributions to o-Ps${}\to3\gamma$ events are
p-Ps${}\to2\gamma$ events with registration of additional scatterings,
partially reconstructed o-Ps${}\to3\gamma$ decays mixed with different hits and
o-Ps${}\to3\gamma$ decays with wrong
assignment of hits to photons from annihilation and scattering.
As an example, a composition of different background events is presented in Fig.~\ref{fig:bcg}.
Differences between signal and background events were  identified in the two-dimensional distribution of
difference and the sum of relative angles between momentum vectors of photons (Fig.~\ref{fig:3D}a for
the full Monte Carlo sample,
Fig.~\ref{fig:3D}b for simulated background events,
and for simulated signal events in the Fig.~\ref{fig:3D}c)~\cite{Moskal:2016moj}.

All the values of the applied cuts were optimized for the best Monte Carlo
to data agreement of the distribution of the sum of two smallest relative azimuthal angles
between the annihilation photons (Fig.~\ref{fig:normalization}).

The number of generated Monte Carlo events exceeds the number of experimental events. 
Therefore,
the normalization of Monte Carlo contributions was performed
with two independent parameters: one for o-Ps${}\to3\gamma$ signal events and a second scaling parameter for remaining background events.
The histograms in Fig.~\ref{fig:normalization} are shown after the normalization procedure.

\begin{figure}[b]
\centering
\includegraphics[width=\textwidth]{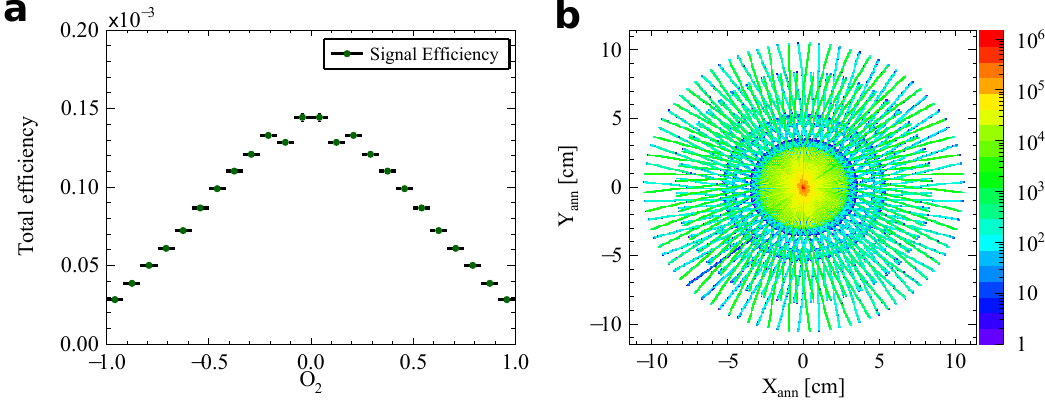}
\caption{The signal efficiency as a function of O$_2$ and the image of $e^+e^-{}\to2\gamma$ annihilation points.
\textbf{a}~Monte Carlo simulation derived distribution of efficiency including all selection criteria applied for the described analysis and
geometrical acceptance of the detector for signal events.
\textbf{b}~Annihilation points in the XY plane reconstructed with $2\gamma$ events from 19 days of measurement.
The visible rosette pattern is due to the geometrical acceptance of the detector
(placement of scintillators strips presented at Fig.~\ref{fig:detector}).
}
\label{fig:effiacc_anniXY}
\end{figure}

The geometrical acceptance of the J-PET detector is determined using Monte Carlo simulations. It is estimated as a ratio of the number of simulated signal events to 
the number of generated events with the o-Ps decay into three photons and one scattered photon. The signal events are those in which three photons from 
the o-Ps to 3$\gamma$ decay interacted in the detector and at least one of them scattered a second time.
The selected Monte Carlo  signal events after the entire analysis chain are used for determination of analysis efficiency as a ratio of
the number of signal events surviving selection cuts over a number of registered signal events. The combined distribution of acceptance and analysis efficiency
is presented 
in the Fig.~\ref{fig:effiacc_anniXY}a.

%%%%%%%%%%%%%%%%%%%%%%%
\subsection*{Determination of expectation value}
Having three hits assigned to an o-Ps${}\to3\gamma$ decay and a fourth hit correlated as a scattering to one of the annihilation photons,
cos$(\omega_{ij})$ is determined for each event after the analysis selection chain. 
The energy of annihilation photons
from o-Ps decay is calculated in the basis of their relative angles~\cite{Kaminska:2016fsn},
and the momenta are ordered accordingly $\mathrm{k}_1 > \mathrm{k}_2 > \mathrm{k}_3$.
Then cos$(\omega_{ij})$ is calculated according to 
Eq.~(\ref{eq:alpha}) while $\upepsilon$ is derived from Eq.~(\ref{eq:epsilon}).
From the experimental distribution of cos$(\omega_{ij})$ (Fig.~\ref{fig:result}b) the normalized background spectrum is subtracted.
The obtained distribution is finally divided by Monte Carlo derived distributions of the total efficiency (Fig.~\ref{fig:effiacc_anniXY}a)
and a mean value of the distribution is calculated as the expectation value of the correlation
$O_2$, Eq.~(\ref{eq:operator}), along with
the statistical error of the expectation value.
%%%%%%%%%%%%%%%%%%%%%%%
\subsection*{Estimation of systematic uncertainties}
Contributions from all selection criteria to the systematic uncertainty were calculated by changing the given cut value by its resolution and performing the whole analysis chain again.
Following the approach proposed by Barlow~\cite{Barlow:2002yb,Barlow:2002bk}
the statistical significance of the systematical contribution from each cut was calculated as
the difference between the expectation value of operator $O_2$ obtained this way and the final result 
normalized to the uncertainty.
The resolution of the distance to the annihilation plane (DOP) is estimated to be 1.1~cm. The position of the energy deposition by a
photon
along the
scintillator strip
Z$_{\text{hit}}$ is known to the accuracy of 2.4~cm. The angular resolution for $\theta_1+\theta_2$ determination (Fig.~\ref{fig:normalization}) is 1.5~deg.
The emission time spread ETS of photons originating from the o-Ps decay is 
about 0.5~ns.
The TOT is measured
with 1.2~ns resolution, while the DAQ registration threshold is known up to 14~keV. 
The position of 
the
annihilation
measured from p-Ps${}\to2\gamma$ decays is known to 0.5~mm
accuracy in the X-Y plane and 0.4~mm resolution along the Z axis.
It is worth to mention 
that the annihilation place (source position) is continuously monitored with p-Ps${}\to2\gamma$ events,
as an intersection of lines formed with two monoenergetic, back-to-back annihilation photons. 
Two dimensional distribution of reconstructed annihilation points in the XY plane is presented
in the Fig.~\ref{fig:effiacc_anniXY}b.
Possible influence of bin width at cos$(\omega_{ij})$ spectrum (Fig.~\ref{fig:result}b) to the final result is tested with
double and twice reduced width of the bin.
Finally the contribution of pure cosmic rays is estimated with a separate measurement without positronium source for which registered data is analyzed the same way as in case
of o-Ps${}\to3\gamma$ decays. The resulting distribution of cos$(\omega_{ij})$ is subtracted from the experimental spectrum.
Additionally, for conservative consideration the cosmic rays distribution is added to the experimental spectrum.
The result shows no statistically significant contribution from any of the aforementioned parameters.

%%%%%%%%%%%%%%%%%%%%%%%%%%%%%%%%%%%%%%%
\section*{Data availability}
The data sets collected in the experiment and analyzed during the current study are available
under restricted access due to the large data volume.
Direct access to the data can be arranged on request
by contacting the corresponding author.

%%%%%%%%%%%%%%%%%%%%%%%%%%%%%%%%%%%%%%%%%%%%%%%%%here goes bibliography

%%%%%%%%%%%%%%%%%%%%%%%%%%%%%%%%%%%%%%%%%%%%%%%%%
\section*{Acknowledgments}
The authors acknowledge the technical and administrative support of A.~Heczko, M.~Kajetanowicz and W.~Migda\l{}.
This work was supported by
the Foundation for Polish Science through the
TEAM POIR.04.04.00-00-4204/17 program (P.M.),
the National Science Centre of Poland through grants MAESTRO no.
2021/42/A/ST2/00423 (P.M.), OPUS no. 2019/35/B/ST2/03562 (P.M.) and SONATA BIS no. 2020/38/E/ST2/00112 (E.~P.d.R.),
the Ministry of Education and Science through grant no. SPUB/SP/490528/2021 (P.M.),
the EU Horizon 2020 research and innovation programme, 
STRONG-2020 project, under grant agreement No 824093 (P.M.),
and the SciMat and qLife Priority Research Areas  budget under the program {\it Excellence Initiative - Research University} at the Jagiellonian University (P.M.),
and Jagiellonian University project no. CRP/0641.221.2020 (P.M.).

%%%%%%%%%%%%%%%%%%%%%%%%%%%%%%%%%%%%%%%%%%%%%%%%%%
\section*{Author contributions}
The experiment was conducted using the J-PET apparatus.
The J-PET detector, the techniques of the experiment, and the symmetry test involving polarization were conceived by P.~M.
The data analysis was conducted by J.~R.
Signal selection criteria were developed by P.~M. and E.~C., applied by J.~R., and verified by E.~C.
{ Authors:
  P.~M.,
  E.~C.,
  J.~R.,
  E.Y.~B.,
  N.~C.,
  A.~C.,
  C.~C.,
  M.~Dadgar,
  M.~Das,
  K.~D.,
  A.~G.,
  B.C.~H.,
  K.~Kacprzak,
  T.~Kaplanoglu,
  \L{}.~K.,
  K.~Klimaszewski,
  P.~K.,
  G.~K.,
  T.~Kozik,
  W.~K.,
  D.~K.,
  S.~M.,
  W.~M.,
  S.~N.,
  S.~P.,
  E.~P.d.R.,
  L.~R.,
  S.~S.,
  S.C.,
  R.Y.~S.,
  M.~Silarski,
  M.~Skurzok,
  E.~\L{}.~S.,
  P.~T.,
  F.~T.A.,
  K.~.T.A.,
  K.~V.E.,
  and
  W.~W.
participated in the construction, commissioning, and operation of the experimental setup, as well as in the data-taking campaign and data interpretation.} 
  M.~G. and B.~J. designed and constructed the positronium production chamber.
  S.~N. and G.~K. optimized the working parameters of the detector.
  K.~D., A.~G., K.~Kacprzak and W.~K. developed the J-PET analysis and simulation framework.
G.~K. developed and operated the DAQ system.
K.~D., M.~Silarski and M.~Skurzok performed timing calibration of the detector.
K.~Klimaszewski, P.~K., W.~W., L.~R. and R.Y.~S. managed the computing resources for high-level analysis and simulations.
E.~C. developed and operated short- and long-term data archiving systems and the computer center of J-PET.
S.~S. established relation between energy loss and TOT.
S.D.~B. provided advice for the theory.
P.~M. managed the whole project and secured the main financing. 
The manuscript was prepared by P.~M., E.~C., S.D.~B., and J.~R. and was then edited and approved by all authors.

%%%%%%%%%%%%%%%%%%%%%%%%%%%%%%%%%%%%%%%%%%%
\section*{Competing interests}
The authors declare no competing interests.
\end{document}